\documentclass{article}
\usepackage{ismir,amsmath,cite}
\usepackage{graphicx}
\usepackage{color}
\usepackage[inline]{enumitem}
\usepackage{amssymb}
\usepackage{hyperref}
\usepackage{adjustbox}
\usepackage{nicefrac}
\usepackage{booktabs}
\usepackage[utf8]{inputenc}
\usepackage[english]{babel}
\usepackage{xspace}

\newcommand{\tS}{\textsuperscript}

\title{Feature Learning for Chord Recognition:\\ The Deep Chroma Extractor}

\oneauthor
{Filip Korzeniowski and Gerhard Widmer}
{Department of Computational Perception, Johannes Kepler University Linz, Austria
\\ {\tt filip.korzeniowski@jku.at}}

\let\OLDthebibliography\thebibliography
\renewcommand\thebibliography[1]{
  \OLDthebibliography{#1}
  \setlength{\parskip}{0pt}
  \setlength{\itemsep}{4.0pt plus 0.3ex}
}

\begin{document}

\maketitle
\begin{abstract}
We explore frame-level audio feature learning for chord recognition using
artificial neural networks. We present the argument that chroma vectors
potentially hold enough information to model harmonic content of audio for
chord recognition, but that standard chroma extractors compute too noisy
features. This leads us to propose a learned chroma feature extractor based on
artificial neural networks. It is trained to compute chroma features that
encode harmonic information important for chord recognition, while being
robust to irrelevant interferences. We achieve this by feeding the network an
audio spectrum with context instead of a single frame as input. This way, the
network can learn to selectively compensate noise and resolve harmonic
ambiguities.

We compare the resulting features to hand-crafted ones by using a simple
linear frame-wise classifier for chord recognition on various data sets. The
results show that the learned feature extractor produces superior chroma
vectors for chord recognition.
\end{abstract}

\section{Introduction}\label{sec:introduction}

Chord Recognition (CR) has been an active research field since its inception by
Fujishima in 1999 \cite{fujishima_realtime_1999}. Since then, researchers have
explored many aspects of this field, and developed various systems to
automatically extract chords from audio recordings of music (see
\cite{mcvicar_automatic_2014} for a recent review). Chord recognition meets
this great interest in the MIR (music information research) community because
harmonic content is a descriptive mid-level feature of (Western) music that can
be used directly (e.g.\ for creating lead sheets for musicians) and as basis
for higher-level tasks such as cover song identification, key detection or
harmonic analysis.

Most chord recognition systems follow a common pipeline of feature extraction,
pattern matching, and chord sequence decoding (also called
\emph{post-filtering}) \cite{cho_relative_2014}. In this paper, we focus on the
first step in this pipeline: feature extraction.

Two observations lead us to explore better features for chord recognition:
\begin{enumerate*}[label=(\arabic*)]
    \item \emph{The capabilities of chord models for pattern matching
        are limited}. In \cite{cho_relative_2014}, Cho and Bello conclude that
        appropriate features largely redeem the benefits of complex chord
        models.
    \item \emph{The capabilities of post-filtering are limited}. As
        shown in \cite{cho_relative_2014,chen_chord_2012}, post-filtering
        methods are useful because they enforce continuity of individual chords
        rather than providing information about chord transitions.
        Incorporating such information did not considerably improve recognition
        results in both studies. Chen et al.\cite{chen_chord_2012} also observed
        quantitatively that in popular music ``chord progressions are less
        predictable than it seems'', and thus knowing chord history does not
        greatly narrow the possibilities for the next chord.
\end{enumerate*}
Given these apparent limitations of the pattern matching and post-filtering
stages, it is not surprising that they only partly compensate the performance
gap between features \cite{cho_relative_2014}. We therefore have to compute better
features if we want to improve chord recognition.

In this paper, we take a step towards better features for chord recognition by
introducing a data-driven approach to extract chromagrams that
specifically encode content relevant to harmony. Our method learns
to discard irrelevant information like percussive noise, overtones or timbral
variations automatically from data. We argue that it is thus able to
compensate a broader range of interferences than hand-crafted approaches.

\section{Chromagrams}
\label{sec:features}

The most popular feature used for chord recognition is the \emph{Chromagram}.
A chromagram comprises a time-series of chroma vectors, which represent
harmonic content at a specific time in the audio as \(c \in \mathbb{R}^{12}\).
Each \(c_i\) stands for a pitch class, and its value indicates the current
saliency of the corresponding pitch class. Chroma vectors are computed by
applying a filter bank to a time-frequency representation of the
audio. This representation results from either a short-time Fourier transform
(STFT) or a constant-q transform (CQT), the latter being more popular due to a
finer frequency resolution in the lower frequency area.

Chromagrams are concise descriptors of harmony because they encode tone quality
and neglect tone height. In theory, this limits their representational
power: without octave information, one cannot distinguish e.g.\ chords
that comprise the same pitch classes, but have a different bass note (like G
vs.\ G/5, or A:sus2 vs.\ E:sus4).
In practice, we can show that given chromagrams derived from ground truth
annotations, using logistic regression we can recognise ~97\% of chords
(reduced to major/minor) in the Beatles dataset. This result encourages us to
create chroma features that contain harmony information, but are robust to
spectral content that is harmonically irrelevant.

Chroma features are noisy in their basic formulation because they are
affected by various interferences: musical instruments produce overtones in
addition to the fundamental frequency; percussive instruments pollute the
spectrogram with broadband frequency activations (e.g.\ snare drums) and/or
pitch-like sounds (tom-toms, bass drums); different combinations of instruments
(and different, possibly genre-dependent mixing techniques) create different
timbres and thus increase variance \cite{cho_relative_2014,
mcvicar_automatic_2014}.

Researchers have developed and used an array of methods that mitigate these
problems and extract cleaner chromagrams: Harmonic-percussive source
separation can filter out broadband frequency responses of percussive
instruments  \cite{ono_separation_2008,ueda_hmm_2010}, various methods
tackle interferences caused by overtones \cite{mauch_approximate_2010,
cho_relative_2014}, while \cite{muller_making_2009,ueda_hmm_2010} attempt
to extract chromas robust to timbre. See \cite{cho_relative_2014} for a recent
overview and evaluation of different methods for chroma extraction. Although
these approaches improve the quality of extracted chromas, it is very
difficult to hand-craft methods for all conceivable disturbances, even if
we could name and quantify them.

The approaches mentioned above share a common limitation: they mostly operate
on single feature frames. Single frames are often not
enough to decide which frequencies salient in the spectrum are relevant to
harmony and which are noise. This is usually countered by contextual
aggregation such as moving mean/median filters or beat synchronisation, which
are supposed to smooth out noisy frames. Since they operate only \emph{after}
computing the chromas, they address the symptoms (noisy frames) but do not
tackle the cause (spectral content irrelevant to harmony). Also,
\cite{cho_relative_2014} found that they blur chord boundaries and
details in a signal and can impair results when combined with more complex
chord models and post-filtering methods.

It is close to impossible to find the rules or formulas that define
harmonic relevance of spectral content manually. We thus resort to the data-driven approach of
deep learning. Deep learning was found to extract strong, hierarchical,
discriminative features \cite{bengio_representation_2013} in many domains.
Deep learning based systems established new state-of-the-art results in computer
vision\footnote{See \url{https://rodrigob.github.io/are_we_there_yet/build/classification_datasets_results.html}
for results on computer vision.}, speech recognition, and MIR tasks such as
beat detection \cite{bock_multi_2014}, tempo estimation
\cite{bock_accurate_2015} or structural segmentation
\cite{ullrich_boundary_2014}.

In this paper, we want to exploit the power of deep neural networks to compute
harmonically relevant chroma features. The proposed chroma extractor learns to
filter harmonically irrelevant spectral content from a  \emph{context of audio
frames}. This way, we circumvent the necessity to temporally smooth the
features by allowing the chroma extractor to use context information
directly. Our method computes cleaner chromagrams while retaining their
advantages of low dimensionality and intuitive interpretation.

\section{Related Work}\label{sec:related_work}

A number of works used neural networks in the context of chord
recognition. Humphrey and Bello \cite{humphrey_rethinking_2012} applied
Convolutional Neural Networks to classify major and
minor chords end-to-end. Boulanger-Lewandowski et
al.\cite{boulanger-lewandowski_audio_2013}, and Sigtia et
al.\cite{sigtia_audio_2015} explored Recurrent Neural Networks as a
post-filtering method, where the former used a deep belief net, the latter a
deep neural network as underlying feature extractor. All these approaches
train their models to directly predict major and minor chords, and following
\cite{bengio_representation_2013}, the hidden layers of these models learn
a hierarchical, discriminative feature representation. However, since the models are trained
to distinguish major/minor chords only, they consider other chord types (such
as seventh, augmented, or suspended) mapped to major/minor as intra-class
variation to be robust against, which will be reflected by the extracted
internal features. These features might thus not be useful to recognise
other chords.

We circumvent this by using chroma templates derived from chords as
distributed (albeit incomplete) representation of chords. Instead of directly
classifying a chord label, the network is required to compute the
\emph{chroma representation} of a chord given the corresponding spectrogram.
We expect the network to learn which saliency in the spectrogram is
responsible for a certain pitch class to be harmonically important, and
compute higher values for the corresponding elements of the output chroma
vector.

Approaches to directly learn a mapping from spectrogram to chroma include
those by İzmirli and Dannenberg \cite{izmirli_understanding_2010} and Chen et
al.\cite{chen_chord_2012}. However, both learn only a linear transformation of
the time-frequency representation, which limits the mapping's expressivity.
Additionally, both base their mapping on a single frame, which comes with the
disadvantages we outlined in the previous section.

In an alternative approach, Humphrey et al. apply deep learning methods to
produce Tonnetz features from a spectrogram \cite{humphrey_learning_2012}.
Using other features than the chromagram is a promising direction, and was also
explored in \cite{chen_chord_2012} for bass notes. Most chord recognition
systems however still use chromas, and more research is necessary to
explore to which degree and under which circumstances Tonnetz features
are favourable.

\section{Method}\label{sec:method}

To construct a robust chroma feature extractor, we use a deep neural network
(DNN). DNNs consist of $L$ hidden layers $h_l$ of $U_l$
computing units. These units compute values based on the results of the
previous layer, such that
\begin{align}
    h_l(x) = \sigma_l\left(W_l \cdot h_{l-1}(x) + b_l\right),
    \label{eqn:hidden_layer}
\end{align}
where \(x\) is the input to the net, \(W_l \in \mathbb{R}^{U_l \times
U_{l-1}}\) and \(b_l \in \mathbb{R}^{U_l}\) are the weights and the bias of the
\(l\)\tS{th} layer respectively, and \(\sigma_l\) is a (usually non-linear)
\emph{activation function} applied point-wise.

We define two additional special layers: an \emph{input layer} that is feeding
values to \(h_1\) as \(h_0(x) = x\), with \(U_0\) being the input's
dimensionality; and an \emph{output layer} \(h_{L+1}\) that takes the same form
as shown in Eq.~\ref{eqn:hidden_layer}, but has a specific semantic purpose:
it represents the output of the network, and thus its dimensionality
\(U_{L+1}\) and activation function \(\sigma_{L+1}\) have to be set
accordingly.\footnote{For example, for a 3-class classification problem one
    would use 3 units in the output layer and a softmax activation function
    such that the network's output can be interpreted as probability
    distribution of classes given the data.}

The weights and biases constitute the model's parameters. They are trained in
a supervised manner by gradient methods and error back-propagation
in order to minimise the loss of the network's output. The loss function
depends on the domain, but is generally some measure of difference between the
current output and the desired output (e.g.\ mean squared error, categorical
cross-entropy, etc.)

In the following, we describe how we compute the input to the DNN,
the concrete DNN architecture and how it was trained.

\subsection{Input Processing}

We compute the time-frequency representation of the signal based on the
magnitude of its STFT \(X\). The STFT gives significantly worse results than
the constant-q transform if used as basis for traditional chroma extractors,
but we found in preliminary experiments that our model is not sensitive to this
phenomenon. We use a frame size of 8192 with a hop size of 4410 at a sample
rate of 44100~Hz. Then, we apply triangular filters to convert the linear
frequency scale of the magnitude spectrogram to a logarithmic one in what we
call the \emph{quarter-tone spectrogram} \(S = F^{\triangle}_{Log} \cdot\lvert
X\rvert\), where \(F^{\triangle}_{Log}\) is the filter bank. The quarter-tone
spectrogram contains only bins corresponding to frequencies between 30~Hz and
5500~Hz and has 24 bins per octave. This results in a dimensionality of
178~bins. Finally, we apply a logarithmic compression such that \(S_{log} =
\log\left(1 + S\right)\), which we will call the \emph{logarithmic quarter-tone
spectrogram}. To be concise, we will refer to \(S_{Log}\) as ``spectrogram''
in rest of this paper.

Our model uses a context window around a target frame as input. Through
systematic experiments on the validation folds (see Sec.\ref{sec:compared_features})
we found a context window of \(\pm 0.7\)~s to work best. Since we operate
at 10~fps, we feed our network at each time 15 consecutive frames, which we
will denote as \emph{super-frame}.

\subsection{Model}

We define the model architecture and set the model's hyper-parameters based on
validation performance in several preliminary experiments. Although a more
systematic approach might reveal better configurations, we found that results
do not vary by much once we reach a certain model complexity.

\begin{figure}
 \centerline{
 \includegraphics[width=0.8\columnwidth]{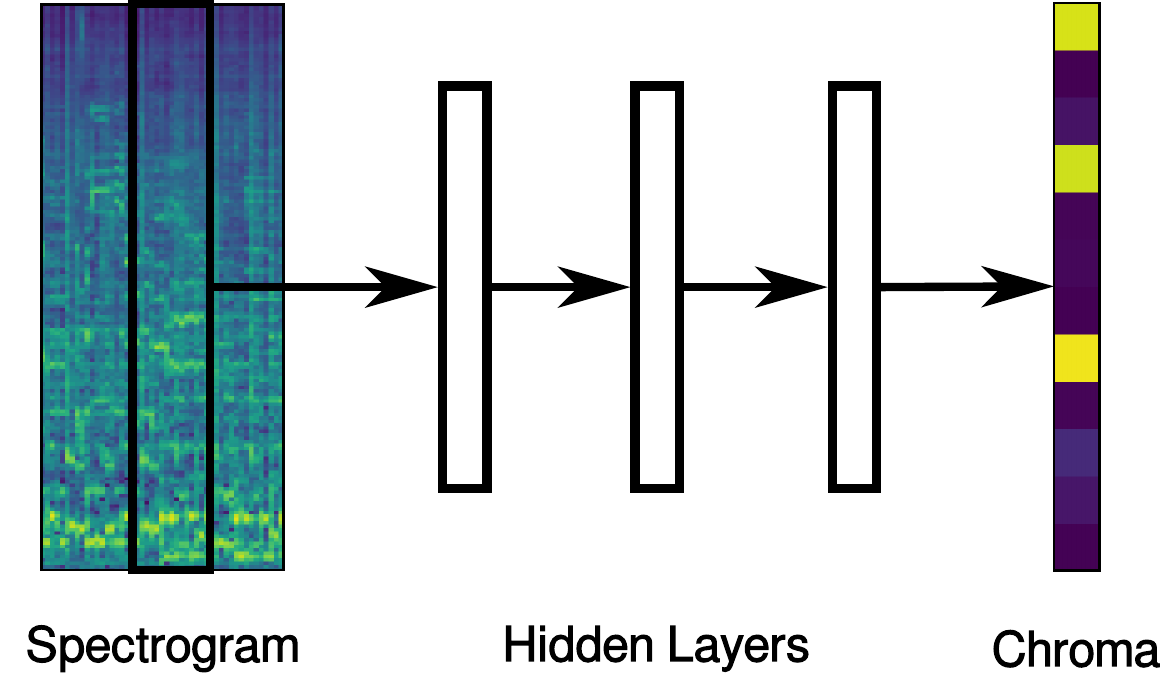}}
 \caption{Model overview. At each time 15 consecutive frames of the input quarter-tone
          spectrogram \(S_{Log}\) are fed to a series of 3 dense layers of 512
          rectifier units, and finally to a sigmoid output layer of 12 units
          (one per pitch class), which represents the chroma vector for the centre
          input frame.}
\label{fig:model}
\end{figure}

Our model is a deep neural network with 3 hidden layers of 512 rectifier
units\cite{glorot_deep_2011} each. Thus, \(\sigma_l(x) = \max(0, x)\) for \(1
\leq l \leq L\). The output layer, representing the chroma vector, consists of
12 units (one unit per pitch class) with a sigmoid activation function
\(\sigma_{L+1}(x) = \nicefrac{1}{1 + \exp(-x)}\). The input layer represents
the input super-frame and thus has a dimensionality of 2670.
Fig.~\ref{fig:model} shows an overview of our model.

\subsection{Training}

To train the network, we propagate back through the network the gradient of
the loss \(\mathcal{L}\) with relation to the network parameters. Our loss is
the binary cross-entropy between each pitch class in the predicted chroma
vector \(\mathbf{p} = h_{L+1}(S_{log})\) and the target chroma vector
\(\mathbf{t}\), which is derived from the ground truth chord label. For a
single data instance,
\begin{align}
    \mathcal{L} = \frac{1}{12} \sum_{i=1}^{12} -t_i \log(p_i) - (1 - t_i) \log(1 - p_i).
\end{align}

We learn the parameters with mini-batch training (batch size 512) using the
ADAM update rule \cite{kingma_adam_2014}. We also tried simple stochastic
gradient descent with Nesterov momentum and a number of manual learn rate
schedules, but could not achieve better results (to the contrary, using ADAM
training usually converged earlier). To prevent over-fitting, we apply
dropout\cite{srivastava_dropout_2014} with probability 0.5 after each hidden
layer and early stopping if validation accuracy does not increase after 20
epochs.

\section{Experiments}
\label{sec:experiments}

To evaluate the chroma features our method produces, we set up a simple chord
recognition task.
We ignore any post-filtering methods and use a simple, linear
classifier (logistic regression) to match features to chords. This way we want
to isolate the effect of the feature on recognition accuracy. As it is common,
we restrict ourselves to distinct only major/minor chords, resulting in 24
chord classes and a 'no chord' class.

Our compound evaluation dataset comprises the Beatles
\cite{harte_automatic_2010}, Queen and Zweieck \cite{mauch_omras2_2009}
datasets (which form the ``Isophonics'' dataset used in the
MIREX\footnote{\url{http://www.music-ir.org/mirex}} competition), the RWC pop
dataset\footnote{Chord annotations available at
\url{https://github.com/tmc323/Chord-Annotations}}\cite{goto_rwc_2002},
and the Robbie Williams dataset\cite{digiorgi_automatic_2013}. The datasets
total 383 songs or approx.\ 21 hours and 39 minutes of music.

We perform 8-fold cross validation with random splits. For the Beatles
dataset, we ensure that each fold has the same album distribution.
For each test fold, we use six of the remaining folds for training and one for
validation.

As evaluation measure, we compute the Weighted Chord Symbol Recall (WCSR),
often called Weighted Average Overlap Ratio (WAOR) of major and minor
chords using the mir\_eval library\cite{raffel_mir_2014}.

\subsection{Compared Features}
\label{sec:compared_features}

We evaluate our extracted features \(C_D\) against three baselines: a standard
chromagram \(C\) computed from a constant-q transform, a chromagram with
frequency weighting and logarithmic compression of the underlying constant-q
transform \(C^W_{Log}\), and the quarter-tone
spectrogram \(S_{Log}\). The chromagrams are computed using the librosa
library\footnote{https://github.com/bmcfee/librosa}. Their parametrisation
follows closely the suggestions in \cite{cho_relative_2014}, where
\(C^W_{Log}\) was found to be the best chroma feature for chord recognition.

\begin{figure}
 \centerline{
 \includegraphics[width=\columnwidth]{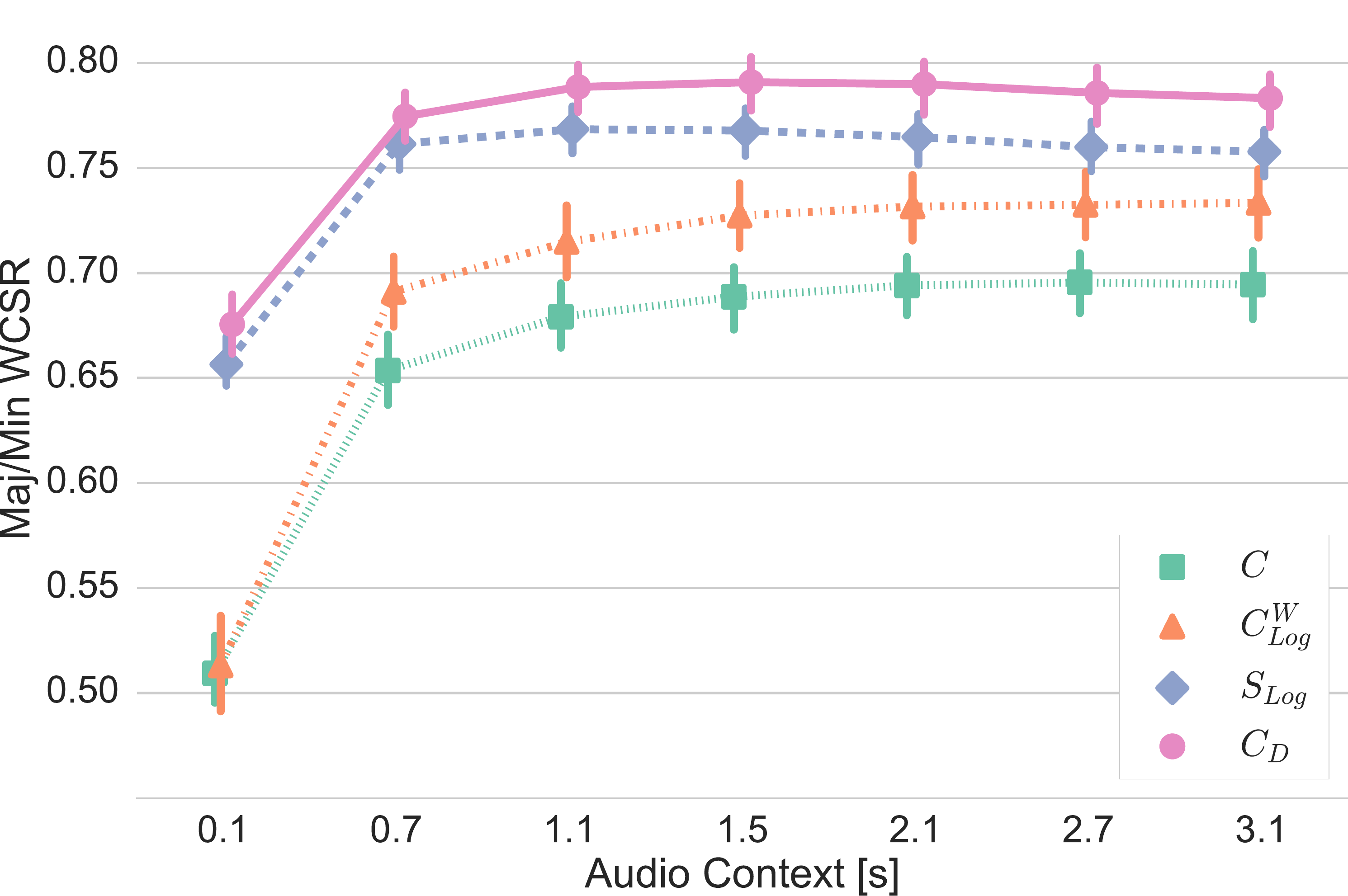}}
 \caption{Validation WCSR for Major/minor chord recognition of different
     methods given different audio context sizes. Whiskers represent 0.95
     confidence intervals.} \label{fig:context}
\end{figure}

Each baseline can take advantage of context information. Instead of computing
a running mean or median, we allow logistic regression to consider multiple
frames of each feature\footnote{Note that this description applies only to the
    baseline methods. For our DNN feature extractor, ``context'' means the
    amount of context the DNN sees. The logistic regression always sees only
one frame of the feature the DNN computed.}.
This is a more general way to incorporate context, because running mean is a
subset of the context aggregation functions possible in our setup. Since
training logistic regression is a convex problem, the result is at least as
good as if we used a running mean.

We determined the optimal amount of context for each baseline experimentally
using the validation folds, as shown in Fig.~\ref{fig:context}. The best
results achieved were 79.0\% with 1.5~s context for \(C_D\), 76.8\% with 1.1~s
context for \(S_{Log}\), 73.3\% with 3.1~s context for \(C^W_{Log}\), and 69.5\%
with 2.7~s context for \(C\). We fix these context lengths for testing.

\section{Results}

\begin{table}[]
\small
\centering
\newcommand{\std}[1]{\xspace{\tiny$\pm$#1}}
\begin{adjustbox}{width=\columnwidth}
\begin{tabular}{@{}rccccc@{}}
\toprule
              & Btls          &  Iso           & RWC           & RW            & Total         \\ \midrule
\(C\)         & 71.0\std{0.1} & 69.5 \std{0.1} & 67.4\std{0.2} & 71.1\std{0.1} & 69.2\std{0.1}          \\
\(C^W_{Log}\) & 76.0\std{0.1} & 74.2 \std{0.1} & 70.3\std{0.3} & 74.4\std{0.2} & 73.0\std{0.1}          \\
\(S_{Log}\)   & 78.0\std{0.2} & 76.5 \std{0.2} & 74.4\std{0.4} & 77.8\std{0.4} & 76.1\std{0.2}          \\
\(C_D\)       & \textbf{80.2}\std{0.1} &  \textbf{79.3}\std{0.1} & \textbf{77.3}\std{0.1} & \textbf{80.1}\std{0.1} & \textbf{78.8}\std{0.1} \\ \bottomrule
\end{tabular}
\end{adjustbox}
\caption{Cross-validated WCSR on the Maj/min task of compared methods on
         various datasets. Best results are bold-faced ($p < 10^{-9}$). Small
         numbers indicate standard deviation over 10 experiments.
         ``Btls'' stands for the Beatles, ``Iso'' for Isophonics, and ``RW'' for the Robbie Williams datasets.
         Note that the Isophonics dataset comprises the Beatles, Queen and
         Zweieck datasets.
     }
\label{tab:results}
\end{table}

Table \ref{tab:results} presents the results of our method compared to the
baselines on several datasets. The chroma features \(C\) and \(C^W_{Log}\)
achieve results comparable to those \cite{cho_relative_2014} reported on a
slightly different compound dataset. Our proposed feature extractor \(C_D\)
clearly performs best, with $p < 10^{-9}$ according to a paired t-test. The
results indicate that the chroma vectors extracted by the proposed method are
better suited for chord recognition than those computed by the baselines.

\begin{figure}[h]
 \centerline{
 \includegraphics[width=\columnwidth]{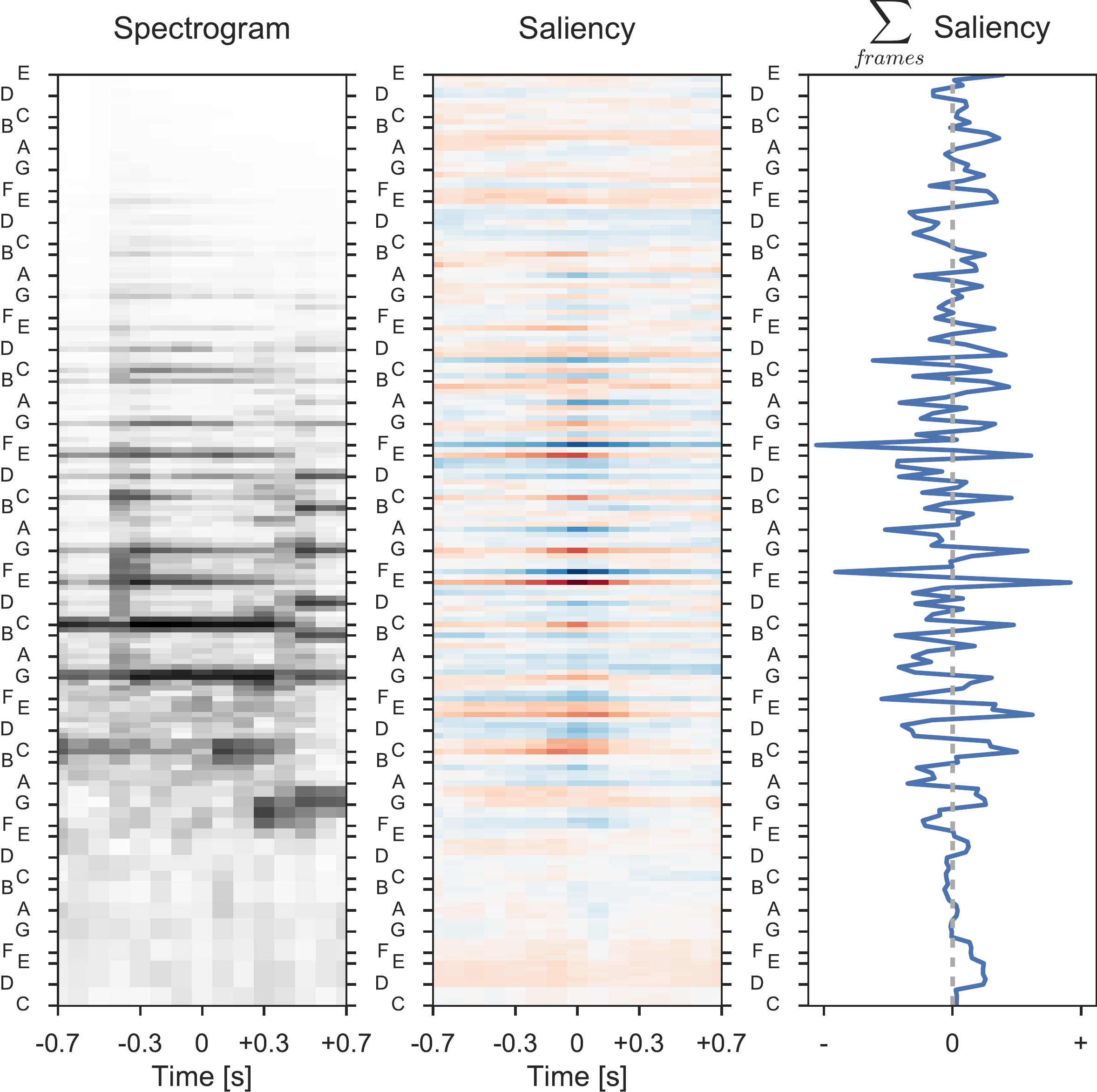}}
 \caption{Input example (C major chord) with corresponding saliency map. The
     \emph{left image} shows the spectrogram frames fed into the network. The
     \emph{centre image} shows the corresponding saliency map, where red pixels
     represent positive, blue pixels negative values. The stronger the
     saturation, the higher the absolute value.
     The \emph{right plot}
     shows the saliency summed over the time axis, and thus how each frequency
     bin influences the output. Note the strong positive influences of frequency
     bins corresponding to \emph{c}, \emph{e}, and \emph{g} notes that form a
     C major chord.}
 \label{fig:saliency_example}
\end{figure}

\begin{figure}[h]
 \centerline{
 \includegraphics[width=\columnwidth]{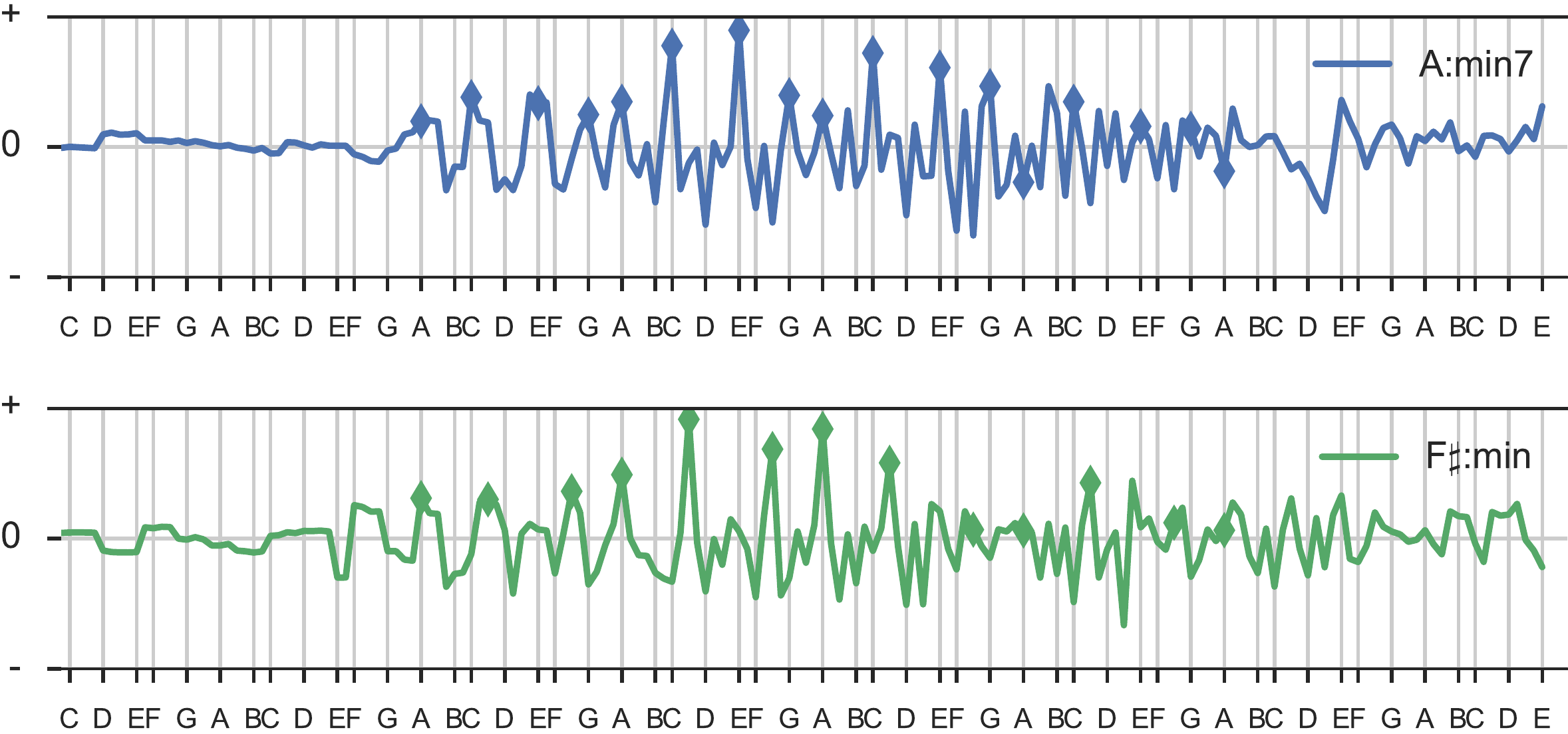}}
 \caption{Average saliency map summed over the time axis for A:min7 and F$\sharp$:min
    chords computed on the Beatles dataset. As expected, we observe mostly
    positive peaks for frequency bins corresponding to notes present in the
    chords (a, c, e, g for A:min7; f$\sharp$, a, c$\sharp$ for F$\sharp$:min).
 }
 \label{fig:chord_saliencies}
\end{figure}

To our surprise, the raw quarter-tone spectrogram \(S_{Log}\) performed better
than the chroma features. This indicates that computing chroma vectors in the
traditional way mixes harmonically relevant features found in the
time-frequency representation with irrelevant ones, and the final classifier
cannot disentangle them. This raises the question of why chroma features are
preferred to spectrograms in the first place. We speculate that the main reason
is their much lower dimensionality and thus ease of modelling (e.g.\ using
Gaussian mixtures).

Artificial neural networks often give good results, but it is difficult to
understand what they learned, or on which basis they generate their output. In
the following, we will try to dissect the proposed model, understand its
workings, and see what it pays attention to. To this end, we compute
saliency maps using guided back-propagation\cite{springenberg_striving_2014},
adapting code freely available\footnote{\url{https://github.com/Lasagne/Recipes/}}
for the Lasagne library\cite{dieleman_lasagne_2015}. Leaving out
the technical details, a saliency map can be interpreted as an attention map
of the same size as the input. The higher the absolute saliency at a specific
input dimension, the stronger its influence on the output, where positive
values indicate a direct relationship, negative values an indirect one.

Fig.~\ref{fig:saliency_example} shows a saliency map and its corresponding
super-frame, representing a C major chord. As expected, the saliency map shows
that the most relevant parts of the input are close to the target frame and
in the mid frequencies. Here, frequency bins corresponding to notes contained
in a C major chord (c, e, and g) showing positive saliency peeks, with the
third, e, standing out as the strongest. Conversely, its neighbouring semitone,
f, exhibits strong negative saliency values. Fig.~\ref{fig:chord_saliencies}
depicts average saliencies for two chords computed over the whole Beatles corpus.

\begin{figure}
 \centerline{
 \includegraphics[width=\columnwidth]{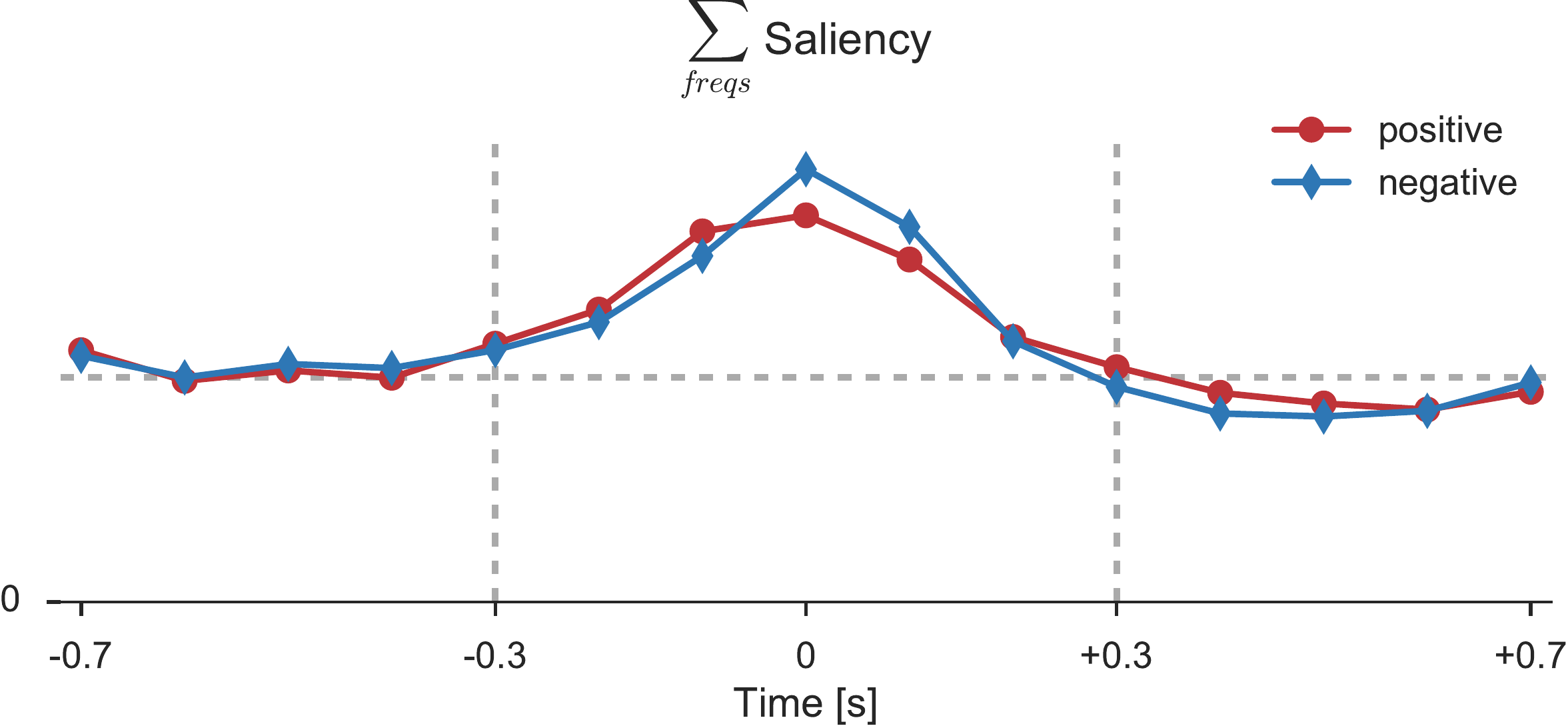}}
 \caption{Average positive and negative saliencies of all input frames of the
          Beatles dataset, summed over the frequency axis. Most of the
          important information is within $\pm 0.3$~s around the centre frame,
          and past data seems to be more important than future data. Around the
          centre frame, the network pays relatively more attention to what
          should be \emph{missing} than present in a given chroma vector, and
          vice versa in areas further away from the centre. The differences are
          statistically significant due to the large number of samples.}
 \label{fig:saliency_total_framewise}
\end{figure}

Fig.~\ref{fig:saliency_total_framewise} shows the average saliency map over
all super-frames of the Beatles dataset summed over the frequency axis. It
thus shows the magnitude with which individual frames in the super-frame
contribute to the output of the neural network. We observe that most
information is drawn from a \(\pm 0.3\)~s window around the centre frame. This
is in line with the results shown in Fig.~\ref{fig:context}, where the
proposed method already performed well with 0.7~s of audio context.

\begin{figure}
 \centerline{
 \includegraphics[width=\columnwidth]{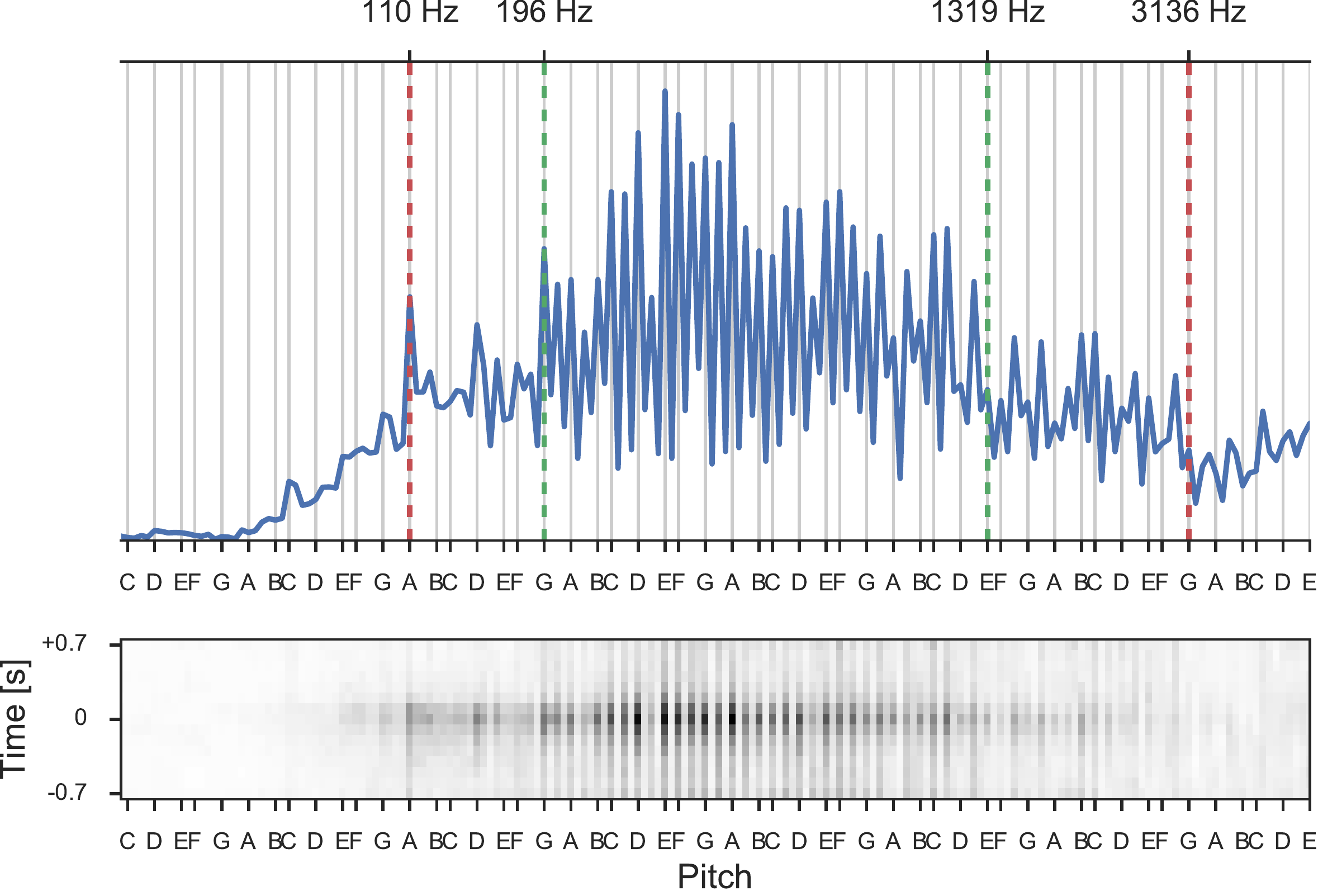}}
 \caption{Average saliency of all input frames of the Beatles dataset (bottom
     image), summed over the time axis (top plot). We see that most relevant
     information can be collected in barely 3 octaves between G3 at 196~Hz
     and E6 at 1319~Hz. Hardly any harmonic information resides below 110~Hz and
     above 3136~Hz. The plot is spiky at frequency bins that correspond to clean
     semitones because most of the songs in the dataset seem to be tuned to a
     reference frequency of 440~Hz. The network thus usually pays little
     attention to the frequency bins between semitones.
 }
 \label{fig:saliency_total_freqwise}
\end{figure}

Fig.~\ref{fig:saliency_total_freqwise} shows the average saliency map over all
super-frames of the Beatles dataset, and its sum over the time axis. We
observe that frequency bins below 110~Hz and above 3136~Hz (wide limits) are
almost irrelevant, and that the net focuses mostly on the frequency range
between 196~Hz and 1319~Hz (narrow limits). In informal experiments, we could
confirm that recognition accuracy drops only marginally if we restrict the
frequency range to the wide limits, but significantly if we restrict it to the
narrow limits. This means that the secondary information captured by the
additional frequency bins of the wide limits is also crucial.

\begin{figure}
 \centerline{
 \includegraphics[width=\columnwidth]{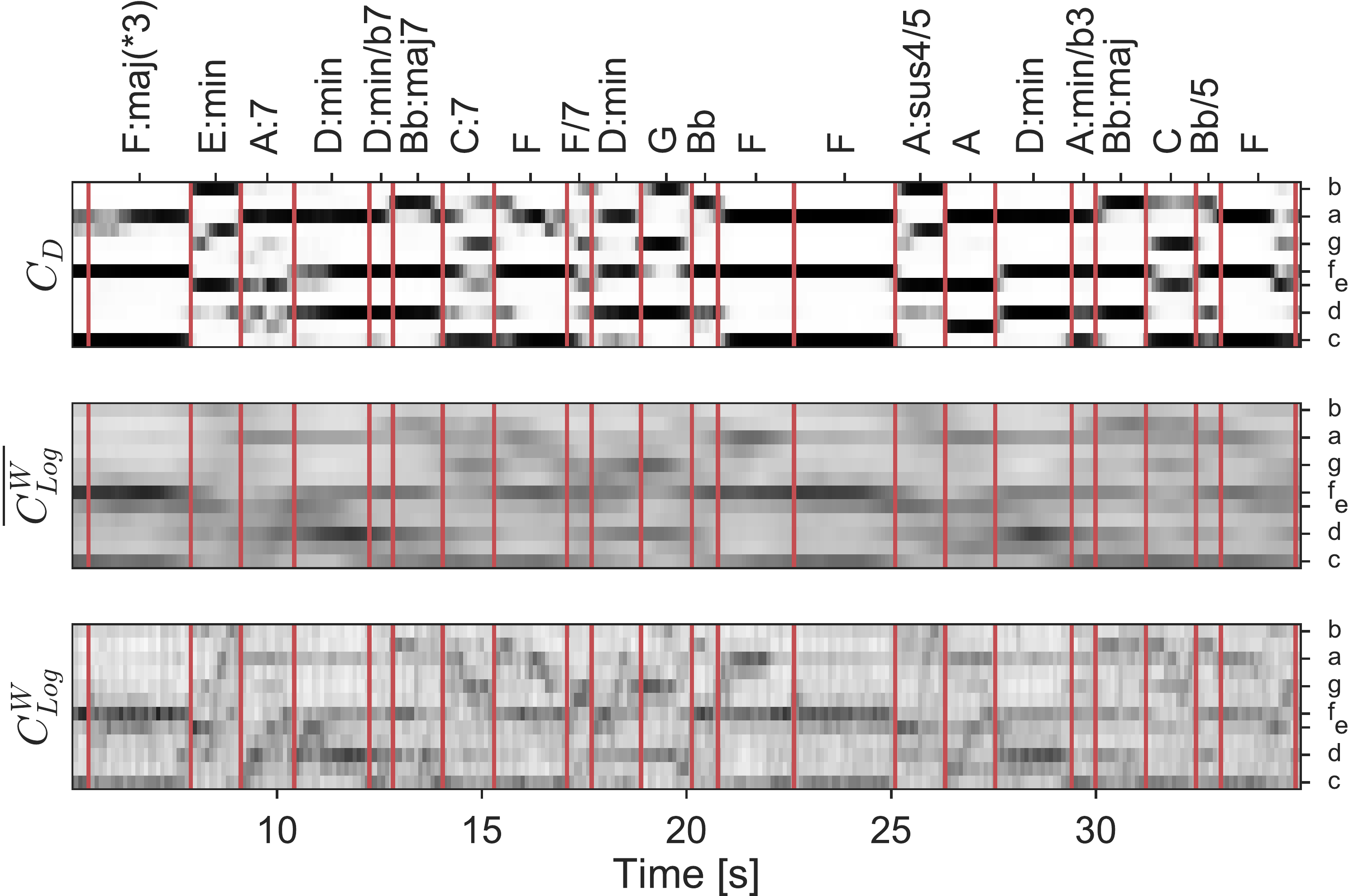}}
 \caption{Excerpts of chromagrams extracted from the song ``Yesterday'' by the
     Beatles.  The \emph{lower image} shows chroma computed by the
     \(C^W_{Log}\) without smoothing. We see a good temporal resolution, but
     also noise. The centre image shows the same chromas after a moving average
     filter of 1.5 seconds. The filter reduced noise considerably, at the cost
     blurring chord transitions. The upper plot shows the chromagram extracted
     by our proposed method. It displays precise pitch activations and low
     noise, while keeping chord boundaries crisp. Pixel values are scaled such
     that for each image, the lowest value in the respective chromagram is
     mapped to white, the highest to black.
 }
 \label{fig:chroma}
\end{figure}

To allow for a visual comparison of the computed features, we depict different
chromagrams for the song ``Yesterday'' by the Beatles in
Fig.~\ref{fig:chroma}. The images show that the chroma vectors extracted by
the proposed method are less noisy and chord transitions are crisper compared
to the baseline methods.

\section{Conclusions and Future Work}

In this paper, we presented a data-driven approach to learning a
neural-network-based chroma extractor for chord recognition.
The proposed extractor computes cleaner chromagrams than state-of-the-art
baseline methods, which we showed quantitatively in a simple chord recognition
experiment and examined qualitatively by visually comparing extracted
chromagrams.

We inspected the learned model using saliency maps and found that a
frequency range of 110~Hz to 3136~Hz seems to suffice as input to
chord recognition methods. Using saliency maps and preliminary
experiments on validation folds we also found that a context of
1.5~seconds is adequate for local harmony estimation.

There are plenty possibilities for future work to extend and/or improve our
method. To achieve better results, we could use DNN ensembles instead of a
single DNN. We could ensure that the network sees data for which its
predictions are wrong more often during training, or similarly, we could
simulate a more balanced dataset by showing the net super-frames of rare
chords more often. To further assess how useful the extracted features are for
chord recognition, we shall investigate how well they interact with
post-filtering methods; since the feature extractor is trained discriminatively,
Conditional Random Fields\cite{lafferty_conditional_2001} would be a natural
choice.

Finally, we believe that the proposed method extracts features that are useful
in any other MIR applications that use chroma features (e.g.\ structural
segmentation, key estimation, cover song detection). To facilitate respective
experiments, we provide source code for our method as part of the \emph{madmom}
audio processing framework \cite{bock_madmom_2016}. Information and source code
to reproduce our experiments can be found at \url{http://www.cp.jku.at/people/korzeniowski/dc}.

\section{Acknowledgements}
This work is supported by the European Research Council (ERC) under the EU's
Horizon 2020 Framework Programme (ERC Grant Agreement number 670035, project
"Con Espressione"). The Tesla K40 used for this research was donated by the
NVIDIA Corporation.

\bibliography{ismir2016}

\begin{thebibliography}{10}

\bibitem{bengio_representation_2013}
Y.~Bengio, A.~Courville, and P.~Vincent.
\newblock Representation {{Learning}}: {{A Review}} and {{New Perspectives}}.
\newblock {\em IEEE Transactions on Pattern Analysis and Machine Intelligence},
  35(8):1798--1828, Aug. 2013.

\bibitem{bock_madmom_2016}
S.~B{\"o}ck, F.~Korzeniowski, J.~Schl{\"u}ter, F.~Krebs, and G.~Widmer.
\newblock madmom: a new {{Python Audio}} and {{Music Signal Processing
  Library}}.
\newblock {\em arXiv preprint arXiv:1605.07008}, 2016.

\bibitem{bock_multi_2014}
S.~B{\"o}ck, F.~Krebs, and G.~Widmer.
\newblock A multi-model approach to beat tracking considering heterogeneous
  music styles.
\newblock In {\em Proceedings of the 15th {{International Society}} for {{Music
  Information Retrieval Conference}} ({{ISMIR}})}, Taipei, Taiwan, 2014.

\bibitem{bock_accurate_2015}
S.~B{\"o}ck, F.~Krebs, and G.~Widmer.
\newblock Accurate tempo estimation based on recurrent neural networks and
  resonating comb filters.
\newblock In {\em Proceedings of the 16th {{International Society}} for {{Music
  Information Retrieval Conference}} ({{ISMIR}})}, M{\'a}laga, Spain, 2015.

\bibitem{boulanger-lewandowski_audio_2013}
N.~Boulanger-Lewandowski, Y.~Bengio, and P.~Vincent.
\newblock Audio chord recognition with recurrent neural networks.
\newblock In {\em Proceedings of the 14th {{International Society}} for {{Music
  Information Retrieval Conference}} ({ISMIR})}, Curitiba, Brazil, 2013.

\bibitem{chen_chord_2012}
R.~Chen, W.~Shen, A.~Srinivasamurthy, and P.~Chordia.
\newblock Chord recognition using duration-explicit hidden {{Markov}} models.
\newblock In {\em Proceedings of the 13th {{International Society}} for {{Music
  Information Retrieval Conference}} ({{ISMIR}})}, Porto, Portugal, 2012.

\bibitem{cho_relative_2014}
T.~Cho and J.~P. Bello.
\newblock On the {{Relative Importance}} of {{Individual Components}} of
  {{Chord Recognition Systems}}.
\newblock {\em IEEE/ACM Transactions on Audio, Speech, and Language
  Processing}, 22(2):477--492, Feb. 2014.

\bibitem{digiorgi_automatic_2013}
B.~{Di Giorgi}, M.~Zanoni, A.~Sarti, and S.~Tubaro.
\newblock Automatic chord recognition based on the probabilistic modeling of
  diatonic modal harmony.
\newblock In {\em Proceedings of the 8th {{International Workshop}} on
  {{Multidimensional Systems}}}, Erlangen, Germany, 2013.

\bibitem{dieleman_lasagne_2015}
S.~Dieleman, J.~Schl{\"u}ter, C.~Raffel, E.~Olson, S.~K. S{\o}nderby, D.~Nouri,
  E.~Battenberg, A.~van~den Oord, et~al.
\newblock Lasagne: First release, 2015.

\bibitem{fujishima_realtime_1999}
T.~Fujishima.
\newblock Realtime {{Chord Recognition}} of {{Musical Sound}}: a {{System Using
  Common Lisp Music}}.
\newblock In {\em Proceedings of the {{International Computer Music
  Conference}} ({{ICMC}})}, Beijing, China, 1999.

\bibitem{glorot_deep_2011}
X.~Glorot, A.~Bordes, and Y.~Bengio.
\newblock Deep sparse rectifier neural networks.
\newblock In {\em Proceedings of the 14th {{International Conference}} on
  {{Artificial Intelligence}} and {{Statistics}} ({{AISTATS}})}, Fort
  Lauderdale, USA, 2011.

\bibitem{goto_rwc_2002}
M.~Goto, H.~Hashiguchi, T.~Nishimura, and R.~Oka.
\newblock {{RWC Music Database}}: {{Popular}}, {{Classical}} and {{Jazz Music
  Databases}}.
\newblock In {\em Proceedings of the 3rd {{International Conference}} on
  {{Music Information Retrieval}} ({{ISMIR}})}, Paris, France, 2002.

\bibitem{harte_automatic_2010}
C.~Harte.
\newblock {\em Towards {{Automatic Extraction}} of {{Harmony Information}} from
  {{Music Signals}}}.
\newblock Dissertation, Department of Electronic Engineering, Queen Mary,
  University of London, London, United Kingdom, 2010.

\bibitem{humphrey_rethinking_2012}
E.~J. Humphrey and J.~P. Bello.
\newblock Rethinking {{Automatic Chord Recognition}} with {{Convolutional
  Neural Networks}}.
\newblock In {\em 11th {{International Conference}} on {{Machine Learning}} and
  {{Applications}} ({{ICMLA}})}, Boca Raton, USA, 2012.

\bibitem{humphrey_learning_2012}
E.~J. Humphrey, T.~Cho, and J.~P. Bello.
\newblock Learning a robust tonnetz-space transform for automatic chord
  recognition.
\newblock In {\em International {{Conference}} on {{Acoustics}}, {{Speech}} and
  {{Signal Processing}} ({{ICASSP}})}, Kyoto, Japan, 2012.

\bibitem{kingma_adam_2014}
D.~Kingma and J.~Ba.
\newblock Adam: {{A}} method for stochastic optimization.
\newblock {\em arXiv preprint arXiv:1412.6980}, 2014.

\bibitem{lafferty_conditional_2001}
J.~D. Lafferty, A.~McCallum, and F.~C.~N. Pereira.
\newblock Conditional {{Random Fields}}: {{Probabilistic Models}} for
  {{Segmenting}} and {{Labeling Sequence Data}}.
\newblock In {\em Proceedings of the 18th {{International Conference}} on
  {{Machine Learning}} ({{ICML}})}, Williamstown, USA, 2001.

\bibitem{mauch_omras2_2009}
M.~Mauch, C.~Cannam, M.~Davies, S.~Dixon, C.~Harte, S.~Kolozali, D.~Tidhar, and
  M.~Sandler.
\newblock {{OMRAS2}} metadata project 2009.
\newblock In {\em Late {{Breaking Demo}} of the 10th {{International
  Conference}} on {{Music Information Retrieval}} ({{ISMIR}})}, Kobe, Japan,
  2009.

\bibitem{mauch_approximate_2010}
M.~Mauch and S.~Dixon.
\newblock Approximate note transcription for the improved identification of
  difficult chords.
\newblock In {\em Proceedings of the 11th {{International Society}} for {{Music
  Information Retrieval Conference}} ({{ISMIR}})}, Utrecht, Netherlands, 2010.

\bibitem{mcvicar_automatic_2014}
M.~McVicar, R.~Santos-Rodriguez, Y.~Ni, and T.~D. Bie.
\newblock Automatic {{Chord Estimation}} from {{Audio}}: {{A Review}} of the
  {{State}} of the {{Art}}.
\newblock {\em IEEE/ACM Transactions on Audio, Speech, and Language
  Processing}, 22(2):556--575, Feb. 2014.

\bibitem{muller_making_2009}
M.~M{\"u}ller, S.~Ewert, and S.~Kreuzer.
\newblock Making chroma features more robust to timbre changes.
\newblock In {\em International {{Conference}} on {{Acoustics}}, {{Speech}} and
  {{Signal Processing}} ({{ICASSP}})}, Taipei, Taiwan, 2009.

\bibitem{ono_separation_2008}
N.~Ono, K.~Miyamoto, J.~{Le Roux}, H.~Kameoka, and S.~Sagayama.
\newblock Separation of a monaural audio signal into harmonic/percussive
  components by complementary diffusion on spectrogram.
\newblock In {\em 16th {{European Signal Processing Conference}}
  ({{EUSIPCO}})}, Lausanne, France, 2008.

\bibitem{raffel_mir_2014}
C.~Raffel, B.~McFee, E.~J. Humphrey, J.~Salamon, O.~Nieto, D.~Liang, and
  D.~P.~W. Ellis.
\newblock mir\_eval: a transparent implementation of common {{MIR}} metrics.
\newblock In {\em Proceedings of the 15th {{International Conference}} on
  {{Music Information Retrieval}} ({{ISMIR}})}, Taipei, Taiwan, 2014.

\bibitem{sigtia_audio_2015}
S.~Sigtia, N.~Boulanger-Lewandowski, and S.~Dixon.
\newblock Audio chord recognition with a hybrid recurrent neural network.
\newblock In {\em 16th {{International Society}} for {{Music Information
  Retrieval Conference}} ({{ISMIR}})}, M{\'a}laga, Spain, 2015.

\bibitem{springenberg_striving_2014}
J.~T. Springenberg, A.~Dosovitskiy, T.~Brox, and M.~Riedmiller.
\newblock Striving for {{Simplicity}}: {{The All Convolutional Net}}.
\newblock {\em arXiv preprint arXiv:1412.6806}, 2014.

\bibitem{srivastava_dropout_2014}
N.~Srivastava, G.~Hinton, A.~Krizhevsky, I.~Sutskever, and R.~Salakhutdinov.
\newblock Dropout: {{A Simple Way}} to {{Prevent Neural Networks}} from
  {{Overfitting}}.
\newblock {\em The Journal of Machine Learning Research}, 15(1):1929--1958,
  2014.

\bibitem{ueda_hmm_2010}
Y.~Ueda, Y.~Uchiyama, T.~Nishimoto, N.~Ono, and S.~Sagayama.
\newblock {{HMM}}-based approach for automatic chord detection using refined
  acoustic features.
\newblock In {\em International {{Conference}} on {{Acoustics Speech}} and
  {{Signal Processing}} ({{ICASSP}})}, Dallas, USA, Mar. 2010.

\bibitem{ullrich_boundary_2014}
K.~Ullrich, J.~Schl{\"u}ter, and T.~Grill.
\newblock Boundary detection in music structure analysis using convolutional
  neural networks.
\newblock In {\em Proceedings of the 15th {{International Society}} for {{Music
  Information Retrieval Conference}} ({{ISMIR}})}, Taipei, Taiwan, 2014.

\bibitem{izmirli_understanding_2010}
{\"O}.~İzmirli and R.~B. Dannenberg.
\newblock Understanding {{Features}} and {{Distance Functions}} for {{Music
  Sequence Alignment}}.
\newblock In {\em Proceedings of the 11th {{International Society}} for {{Music
  Information Retrieval Conference}} ({{ISMIR}})}, Utrecht, Netherlands, 2010.

\end{thebibliography}

\end{document}